\PassOptionsToPackage{prologue,table}{xcolor}
\documentclass[sigconf]{acmart}

\usepackage{usual}

\newcommand{\para}[1]{\vspace{0.2cm}\noindent\textbf{#1}\hspace{.1cm}}

\AtBeginDocument{%
  }

\copyrightyear{2026}
\acmYear{2026}
\setcopyright{cc}
\setcctype{by}
\acmConference[ICSE '26]{2026 IEEE/ACM 48th International Conference on Software Engineering}{April 12--18, 2026}{Rio de Janeiro, Brazil}
\acmBooktitle{2026 IEEE/ACM 48th International Conference on Software Engineering (ICSE '26), April 12--18, 2026, Rio de Janeiro, Brazil}
\acmPrice{}
\acmDOI{10.1145/3744916.3773102}
\acmISBN{979-8-4007-2025-3/2026/04}

\newcommand{\pname}{\textsc{Locus}\xspace}

\definecolor{darkblue}{rgb}{0, 0, 0.5}
\definecolor{darkgreen}{HTML}{027148}
\definecolor{shadecolor}{HTML}{DCDCDC}
\definecolor{darkred}{HTML}{8b0000}
\definecolor{canary-yellow}{HTML}{EEE7AC}

\definecolor{annotation}{HTML}{C3291C}
\newcommand{\redcircle}[1]{%
  \tikz[baseline=(char.base)]{
    \node[shape=circle, fill=black, text=white, inner sep=1pt] (char) {#1};}%
  \ignorespaces
}

\hypersetup{colorlinks=true, citecolor=darkblue, linkcolor=darkblue, urlcolor=darkblue}

\definecolor{deepgreen}{RGB}{0, 150, 0}

\newcommand{\gc}[1]{\cellcolor{deepgreen!#1}}

\newcommand{\sig}[1]{\textbf{#1}}          %
\newcommand{\insig}[1]{\textcolor{gray}{#1}}

\renewcommand{\sig}[1]{\textbf{\textcolor{black}{#1}}} 

\renewcommand{\insig}[1]{#1}

\newcolumntype{Y}{>{\raggedright\arraybackslash}X}

\crefname{item}{}{}
\Crefname{item}{}{}

\begin{document}

\title{\pname: Agentic Predicate Synthesis for Directed Fuzzing}

\author{Jie Zhu}
\affiliation{%
  \institution{University of Chicago}
  \city{Chicago}
  \country{USA}
}

\author{Chihao Shen}
\affiliation{%
  \institution{University of Maryland}
  \city{College Park}
  \country{USA}
}

\author{Ziyang Li}
\affiliation{%
  \institution{Johns Hopkins University}
  \city{Baltimore}
  \country{USA}
}

\author{Jiahao Yu}
\affiliation{%
 \institution{Northwestern University}
 \city{Evanston}
 \country{USA}
}

\author{Yizheng Chen}
\affiliation{%
  \institution{University of Maryland}
  \city{College Park}
  \country{USA}}

\author{Kexin Pei}
\affiliation{%
  \institution{University of Chicago}
  \city{Chicago}
  \country{USA}
}

\renewcommand{\shortauthors}{Jie Zhu, Chihao Shen, Ziyang Li, Jiahao Yu, Yizheng Chen, Kexin Pei}

\begin{abstract}
Directed fuzzing aims to find program inputs that lead to specified target program states.
It has broad applications, such as debugging system crashes, confirming reported bugs, and generating exploits for potential vulnerabilities.
This task is inherently challenging because target states are often deeply nested in the program, while the search space manifested by numerous possible program inputs is prohibitively large.
Existing approaches rely on branch distances or manually-specified constraints to guide the search; however, the branches alone are often insufficient to precisely characterize progress toward reaching the target states, while the manually specified constraints are often tailored for specific bug types and thus difficult to generalize to diverse target states and programs.

We present \pname, a novel framework to improve the efficiency of directed fuzzing.
Our key insight is to synthesize predicates to capture fuzzing progress as semantically meaningful intermediate states, serving as milestones towards reaching the target states.
When used to instrument the program under fuzzing, they can reject executions unlikely to reach the target states, while providing additional coverage guidance.
To automate this task and generalize to diverse programs, \pname features an agentic framework with program analysis tools to synthesize and iteratively refine the candidate predicates, while ensuring the predicates strictly relax the target states to prevent false rejections via symbolic execution.
Our evaluation shows that \pname substantially improves the efficiency of eight state-of-the-art fuzzers in discovering real-world vulnerabilities, achieving an average speedup of 41.6$\times$.
So far, \pname has found nine previously unpatched bugs, with three already acknowledged with draft patches.

\end{abstract}



\maketitle

\section{Introduction}
\label{sec:introduction}

Directed Grey-box Fuzzing (DGF) aims to search for program inputs leading its execution to reach specific target program states, \eg an index variable used to access an array is out of the array bounds, thereby uncovering potential bugs or vulnerabilities.
It is widely used in software engineering and security applications, including debugging system crashes\cite{tan2023syzdirect, wang2021syzvegas}, testing patches\cite{bohme2013regression, marinescu2013katch}, verifying bug reports from static analysis\cite{christakis2016guiding, murali2024fuzzslice}, and generating Proof-of-Concept (PoC) exploits of vulnerabilities\cite{brumley2008automatic, liu2024oss-fuzz-gen}.
While the original purpose of directed fuzzing is largely not for discovering new bugs, \ie it needs a specified target, it has found numerous impactful zero-day vulnerabilities~\cite{liang2020sequence, nguyen2020binary, wang2025patchfuzz, cao2023oddfuzz, shah2022mc2}.

DGF is challenging, as the target program states are often deeply nested in the program, while the search space introduced by the complexity of real-world software is prohibitively large.
To speed up the search and schedule promising inputs, most existing techniques rely on metrics based on control flow proximity, \eg the distance to target location in the control flow graph, or heuristics based on the semantics of the branch predicates\cite{chen2018angora, li2017steelix, shah2022mc2}, \eg the target state \texttt{if x==42} has the distance metric of $|x-42|$.
However, such feedback is sometimes too sparse or indirect to reliably measure the progress, especially when there is a long chain of \emph{implicit} preconditions guarding the target program states~\cite{mitropoulos2023syntax, holler2012fuzzing, dutra2023formatfuzzer, rigger2020finding, klees2018evaluating, zhang2023automata, kim2023dafl, sherman2025ogharn, aschermann2020ijon}.
For example, triggering CVE-2018-13785 in \texttt{libpng} requires a PNG file to satisfy a precise sequence of preconditions, \ie valid signature, correct chunk layout, specific IHDR fields (\eg bit depth, color type), and a magic image width (\texttt{0x55555555}), to trigger an integer overflow~\cite{kim2023dafl}, while the predicates to explicitly check these conditions are largely absent in the code to provide an incremental progress guidance.

To capture the intricate feedback to improve search efficiency, more advanced approaches identify progress-capturing constraints in the program to drive execution towards satisfying specific temporal orders and preconditions~\cite{meng2022ltl-fuzzer, holler2012fuzzing, lee2021cafl, liang2020sequence, li2017steelix, fioraldi2021invscov, padhye2019fuzzfactory, ispoglou2020fuzzgen, balzarotti2021use, aschermann2020ijon}.
However, these constraints are often manually crafted by experts and tailored to specific target state types, \eg focusing on a temporal memory safety bug like a use-after-free by enforcing an allocate–free–use sequence.
As different programs can have diverse target states and disparate functionalities, the feedback metric effective in one case may not generalize to another~\cite{chen2019enfuzz}.

As Machine Learning (ML) and Large Language Models (LLMs) have demonstrated surprising code reasoning capabilities, there has been a growing interest in extending such capabilities to help guide fuzzing~\cite{patra2016learning}. 
A common approach is to employ LLMs to directly generate inputs or grammar-aware input generators (\ie fuzz driver or harness)~\cite{liu2024oss-fuzz-gen, huang2024halo, liu2024evaluating, shi2024harnessing, zhang2024effective, lyu2024promptfuzz, chen2025elfuzz, yanghybrid25}, where the preconditions for reaching target states are expressed as part of the input grammar constraints. 
However, not all conditions to reach target states can be easily represented as input grammars.
For example, \Cref{fig:motivating-example} shows that an intermediate state (\texttt{!found\_plte}) necessary to trigger the buffer overflow in \texttt{libpng} only emerges in the middle of the execution and cannot be easily checked at the input level.
More importantly, such a task formulation is particularly challenging for LLMs, as \emph{reasoning from the target program states all the way to the input} often requires an extremely long context and convoluted analysis chain~\cite{jiang2024towards, li2025hitchhiker, ding2024vulnerability, li2024llm, risse2025top, risse2024uncovering, li2024enhancing, risse2025top}, a common pitfall for LLM hallucination~\cite{sahoo2024comprehensive, yao2025reasoning}, let alone the challenge to verify the correctness of the LLM generated inputs and harnesses, which can significantly impede the fuzzing progress if the generated constraints are incorrect~\cite{padhye2019fuzzfactory, ispoglou2020fuzzgen}.

\para{Our approach}
We present \pname, a new framework that integrates LLMs' code reasoning capabilities for directed fuzzing by synthesizing semantically meaningful and \emph{verifiable} predicates to guide the search. 
Unlike existing LLM-based approaches that focus on constraining the search space at the input (harness) level, which poses a high reasoning burden on LLMs and can be hard to verify, \pname generalizes the constraint generation to be at arbitrary program points.
Specifically, given a target state to reach, \pname \emph{automates} the analysis about the intermediate program states that verifiably \emph{relax} the target states and synthesizes predicates as a curriculum to capture the gradual progress towards reaching them.
Such predicates serve as the preconditions dominating all executions to reach the target states, providing fine-grained progress feedback to guide the fuzzers for input scheduling and early termination\cite{luo2023selectfuzz, huang2022beacon, srivastava2022sievefuzz}.
As these predicates are implemented as source-level instrumentation, they are \emph{agnostic} to any fuzzer implementations and can thus be integrated without any customization.
Moreover, as the instrumentation is a one-time offline process, the cost of running \pname can be amortized in all the succeeding fuzzing campaigns.

\para{Agentic design}
Automating the generation of effective progress-capturing predicates for diverse target states involves nontrivial reasoning across the entire codebase, spanning multiple functions and their associated data and control flows.
Simple prompting techniques, such as in-context learning, chain-of-thought (CoT) prompting, retrieval-augmented generation, etc., can hardly support sophisticated analysis at the repository level. 
To this end, \pname features an agentic synthesizer-validator workflow, equipped with diverse program analysis tools to support the traversal of the control flow graph, tracking data dependencies, retrieving function calls, and symbolic execution.
They serve as the agent's action space, allowing the LLM to \emph{reason} via CoT and \emph{act} by calling them \cite{yao2023react} to iteratively propose, refine, and validate candidate predicates.

Importantly, by constraining the output space of the agent at the \emph{predicate}-level, \pname ensures any (inevitable) LLM errors are corrected before deployment for fuzzing.
Specifically, \pname's output is validated by both the compiler and symbolic execution~\cite{cadar2008klee}.
The former checks the syntactic correctness, while the latter ensures the generated predicates strictly relax the target states, \ie by checking whether there exists a path that violates the predicates while satisfying the target states.
Such a strict relaxation ensures that the fuzzing execution can be safely early-terminated if it violates the predicate.
Figure~\ref{fig:overview} shows the \pname's workflow.

\para{Results}
We evaluate \pname on the Magma benchmark \cite{hazimeh2020magma} with eight widely used libraries and ten vulnerability types across eight state-of-the-art fuzzers, covering both directed and undirected ones.
\pname achieves a significant 70.3$\times$ speedup on average for directed fuzzers, with up to 214.2$\times$ speedup when integrated to accelerate SelectFuzz\cite{luo2023selectfuzz}, one of the state-of-the-art directed fuzzers.
For coverage-guided fuzzers, \pname accelerates them by 13$\times$ on average, including 15.3$\times$ speedup for the extensively optimized fuzzer like AFL++.
So far, \pname has found nine previously unpatched bugs.
We have responsibly reported all bugs to the maintainers, and three of which already have pending fixes.

\section{Overview}
\label{sec:overview}

\begin{figure*}
     \centering
     \hfill
     \begin{subfigure}[t]{0.2\textwidth}
         \centering
         \includegraphics[height=4cm]{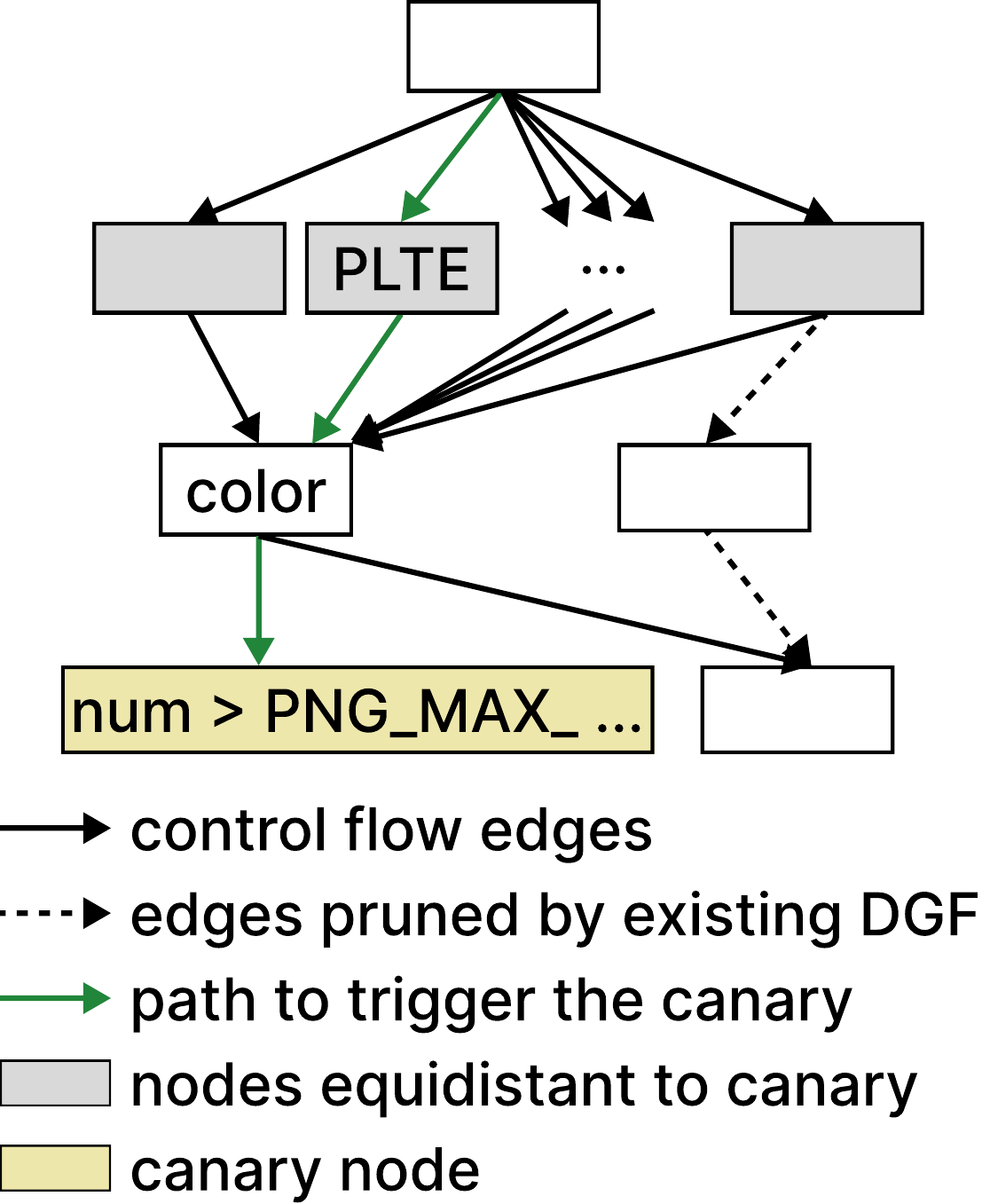}
         \caption{A simplified CFG 
         }
         \label{fig:simplified-cfg}
     \end{subfigure}
     \hfill
     \begin{subfigure}[t]{0.22\textwidth}
         \centering
         \includegraphics[height=4cm]{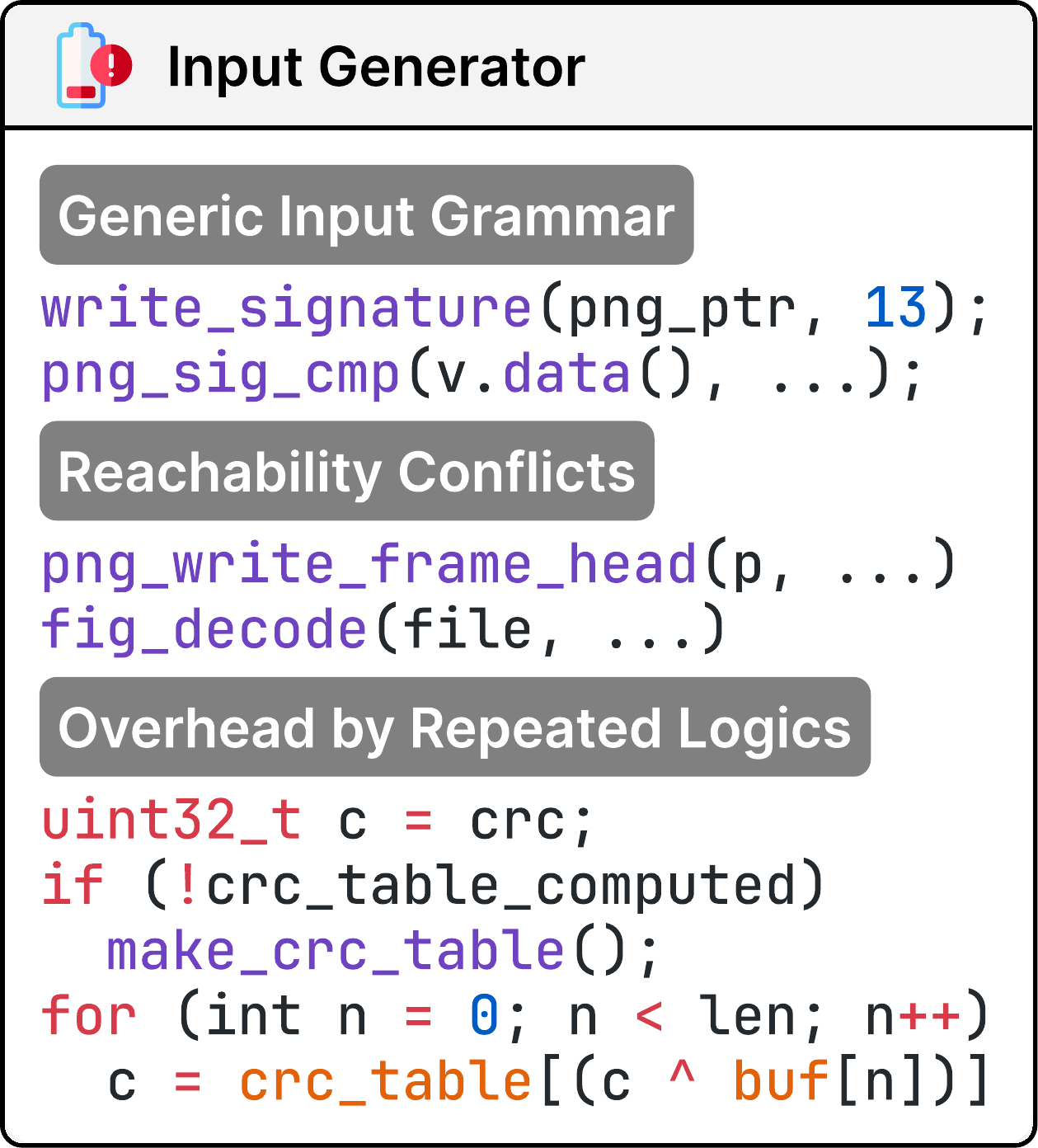}
         \caption{LLM-generated code
         }
         \label{fig:llm-harness}
     \end{subfigure}
     \hfill
     \begin{subfigure}[t]{0.52\textwidth}
         \centering
         \includegraphics[height=4cm]{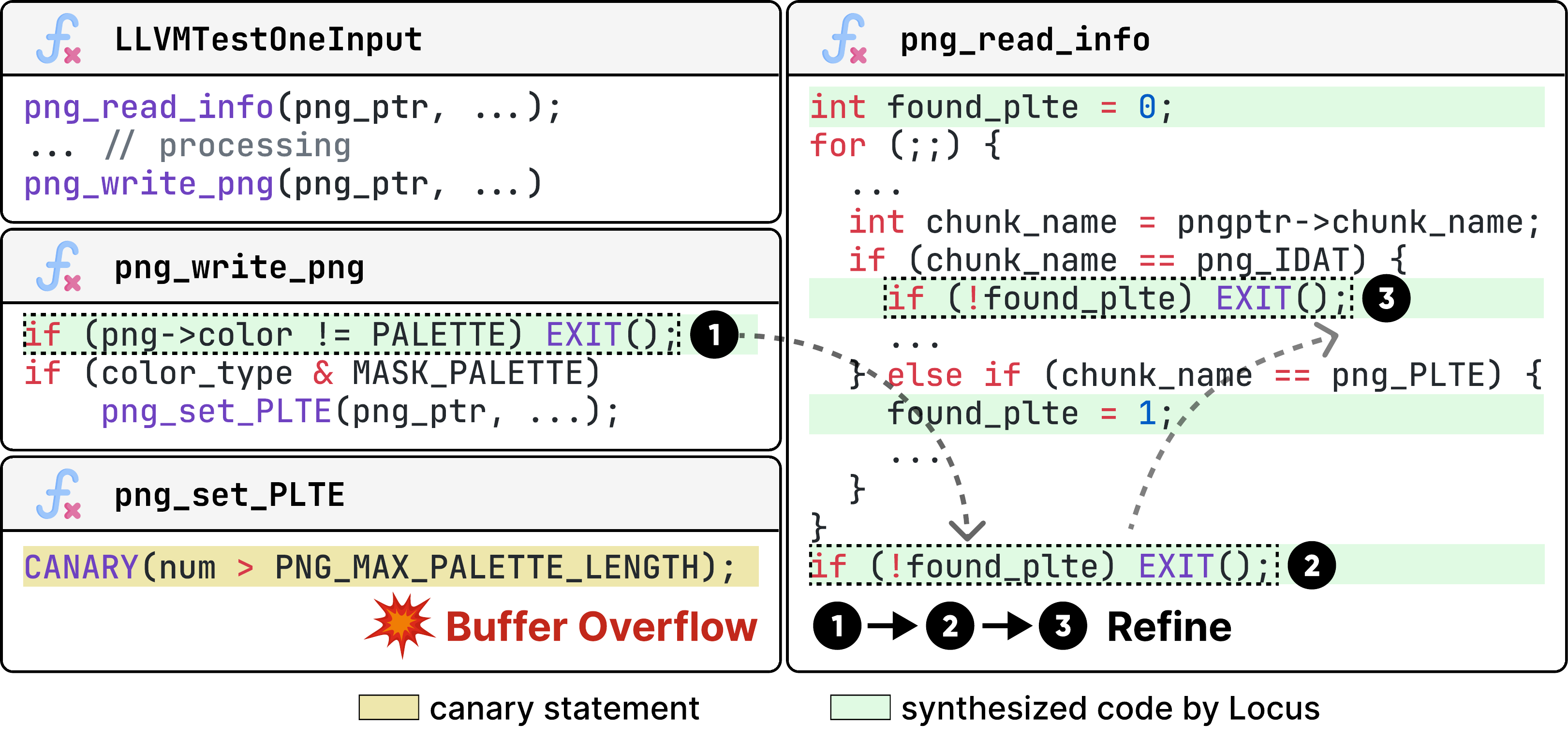}
         \caption{\pname synthesizes predicates in relevant functions. 
         }
         \label{fig:relevant-code}
     \end{subfigure}
        \caption{A motivating example (CVE-2013-6954) showing how \pname complements existing works.
        (a) Traditional approaches based on distance to targets in CFG lack fine-grained guidance to distinguish nodes when they have the same distance.
        (b) LLM-based harness generation is limited to help reach the target.
        (c) Predicates (as \texttt{if} statements) synthesized by \pname provide extra semantic guidance for DGFs, while relaxing the constraint generation from input-level to arbitrary program points.
        }
        \label{fig:motivating-example}
\end{figure*}

In this section, we first describe the background of directed fuzzing.
We then contrast our idea with the existing approaches to demonstrate how \pname complements the existing design (\Cref{fig:motivating-example}).

\subsection{Directed Fuzzing}
\label{subsec:directed-fuzzing}

\para{Canary}
Directed fuzzing aims to generate inputs that drive the program execution to reach predefined program states.
In this paper, we represent these states with \emph{canaries}~\cite{hazimeh2020magma}, which are considered reached when the corresponding canary condition is satisfied.
Early works like AFLGo\cite{bohme2017aflgo} and Beacon\cite{huang2022beacon} treat canaries primarily as the reachability to particular program points, \ie specific line numbers in a code file.
Recent works \cite{hazimeh2020magma,weissberg2024where-to-fuzz} adopt assertion-based canaries, where predicates are explicitly inserted to check vulnerability-triggering conditions at runtime.
Two examples of canaries are highlighted in yellow in \Cref{fig:motivating-example}.

Canaries for directed fuzzing originate from various sources, including static analysis alerts, manually identified vulnerability sites, or runtime sanitizers.
For example, address sanitizers\cite{serebryany2012addresssanitizer} can also be approximately viewed as canaries checking memory safety violations, \eg inserting \texttt{canary(index > maxbound)} to detect out-of-bound accesses.

\para{Common strategies to reach canaries}
Previous research on directed fuzzing can be broadly grouped into four main strategies:
\begin{itemize}[leftmargin=2em]
\item Distance-guided scheduling.
This approach approximates the distance from the currently covered regions of the CFG to the canary, and then prioritizes seeds that are more likely to drive execution along paths that reduce this distance\cite{bohme2017aflgo, du2022windranger,chen2018hawkeye}. 
For paths that are unlikely to reach the canary, the execution can be early terminated\cite{luo2023selectfuzz,huang2022beacon,srivastava2022sievefuzz}.
\item Specialized progress-capturing state representations.
Some approaches introduce domain-specific abstractions tailored for certain classes of vulnerabilities \cite{lee2021cafl, meng2022ltl-fuzzer}.
For example, tracking dataflow from memory allocation to deallocation events for the use-after-free vulnerability.
\item LLM-assisted harness generation.
A more recent direction leverages LLMs to synthesize input generators to manipulate raw program inputs \cite{liu2024inputblaster,xu2025hgfuzzer, zhang2024how, yanghybrid25}.
These methods attempt to constrain the valid execution at the input level by leveraging the learned input grammar knowledge in the LLMs.
\end{itemize}

\subsection{Motivating Example}
\label{subsec:motivating-example}
We use a real-world vulnerability CVE-2013-6954 in \texttt{libpng}, a widely used C library for parsing PNG files, to demonstrate how \pname complements existing approaches.

A simplified CFG of this vulnerability is shown in \Cref{fig:simplified-cfg}, where all the nodes highlighted in gray are equidistant to the canary node (highlighted in yellow), but only one node \texttt{PLTE} is in the path that can lead to the vulnerability (annotated as the green arrows).
Based on this, fuzzers can only perform conservative pruning (annotated as dashed arrows) while omitting a large part of the paths that are irrelevant to the vulnerability, \eg all solid black arrows.
Capturing the progress towards reaching this canary requires specialized knowledge about \texttt{libpng} parsing logic.
However, developing such specialized progress-capturing state representations requires manual efforts from experts and cannot easily generalize to different vulnerabilities and programs.

\begin{figure*}[!t]
\centering
\includegraphics[width=.95\linewidth]{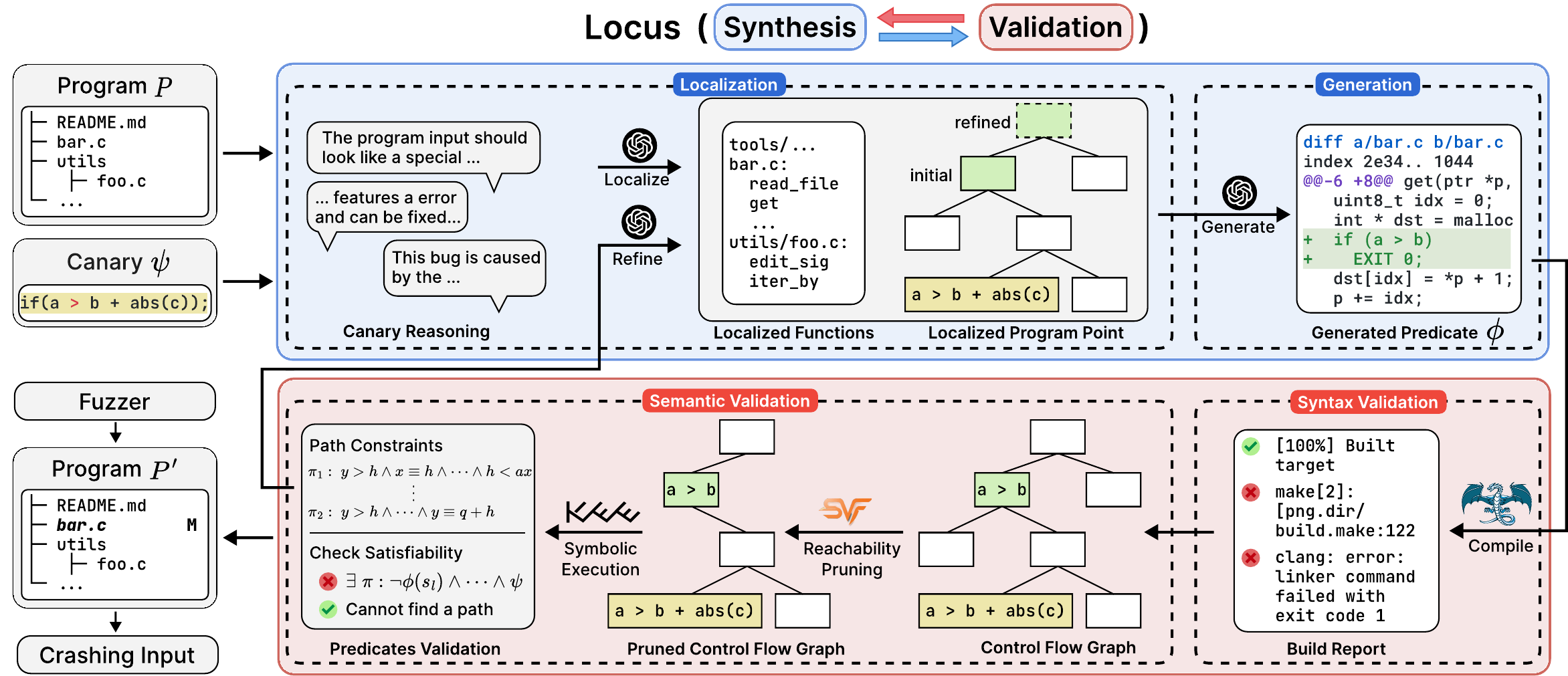}
\caption{
Overview of \pname workflow.
\pname takes as inputs the program codebase \(P\) and the canary \(\psi\), and produces a program \(P'\) instrumented with the progress-capturing predicates. The predicate branches provide extra coverage feedback and guards (via early termination) to guide the fuzzer toward reaching the target state, \ie canary \(\psi\), more efficiently.
}
\label{fig:overview}
\end{figure*}

\Cref{fig:llm-harness} shows an example of the input generator synthesized by LLM-based approaches that enforced some specific input constraints.
However, this approach is inherently limited as program inputs can have sophisticated structures that cannot be easily enforced at the input level, \eg the PNG file contains a compressed data chunk \texttt{IDAT}.
Therefore, the constraints on the input generated by the LLM often reduce to generic input grammars, \eg \texttt{png\_sig\_cmp} only trivially ensures the input is a valid PNG file.
We also note that the synthesized generators cannot be easily checked to determine whether they can effectively help reach the canary.
For example, \texttt{png\_write\_frame\_head} constrains the generated input to a special PNG type \texttt{APNG} that is essentially impossible to reach the canary, but automatically checking this fact is challenging. 
Moreover, to constrain the input to satisfy certain properties necessary to reach the target, the generator sometimes needs to repeat the input processing logic presented in the execution path.
Such repeated execution introduces additional overhead.

\pname synthesizes progress-capturing predicates at arbitrary program points to provide more fine-grained guidance and complement the above approaches.
Specifically, \pname generates a predicate in \texttt{png\_read\_info} (shown in \Cref{fig:relevant-code}), representing the precondition to trigger a real-world buffer overflow vulnerability and terminate the execution if it is not satisfied (\texttt{if(!found\_plte)}).
The target state is highlighted in a \texttt{canary} statement.
\pname's trajectory reveals that it relies on the semantic reasoning of \texttt{libpng}'s parsing behaviors for generating this semantically meaningful predicate.

Specifically, PNG files consist of a series of structured chunks, many of which are optional and can appear in varying orders.
One such optional chunk, \texttt{PLTE}, stores the palette data used for indexed-color images and must adhere to strict size constraints.
In this example, an input PNG file can trigger a buffer overflow in the \texttt{png\_set\_PLTE} function only when the PNG file contains a \texttt{PLTE} chunk, which serves as the necessary state before the target state can be reached, \ie the palette size exceeds the expected bounds.

Existing fuzzers struggle with such a vulnerability due to the complex parsing logic.
Since PNG chunks may appear in arbitrary order and many are optional, the primary parsing routine implements a loop in \texttt{png\_read\_info} to iteratively process each chunk.
Within this loop, multiple branches exist, and each is responsible for handling a specific chunk type, \eg \texttt{IHDR}, \texttt{IDAT}, \texttt{PLTE}.
This loop makes it possible to identify whether a given PNG file contains a \texttt{PLTE} chunk.
However, since these branches are syntactically parallel and executed without a fixed order, they all appear equidistant from the vulnerability site (\texttt{png\_set\_PLTE}) in the control flow graph.
As a result, the naive path distances cannot distinguish between them to prioritize one path over the other.
This explains why, in \Cref{tab:main}, all existing fuzzers take a substantial amount of time to reach this target state in \texttt{libpng}.

\pname starts with generating a predicate to check whether the input PNG file contains a PLTE chunk at the caller of the canary function (\redcircle{1}).
This implies that if the PNG file does not contain a PLTE chunk, the execution can terminate, as it is impossible to reach the target state.
While this predicate is semantically correct, \ie the symbolic execution confirms there is no feasible path to satisfy the target state while violating the generated predicate (\Cref{subsec:validation}), and can help filter out non-palette-based PNG files, it is almost redundant as there is an existing one immediately after the generated check (\texttt{if (color\_type \& MASK\_PALETTE)}), and thus the generated predicate could barely help the fuzzer.

To address this issue, \pname includes another refinement iteration (\Cref{alg:workflow}) to propagate this predicate to a location closer to the program entry such that the infeasible execution can be terminated earlier.
By traversing the call graph and retrieving the functions along the call chain, the refinement proposes a new location in function \texttt{png\_read\_info} right after the parsing loop (\redcircle{2}).
Such a propagated predicates will also be validated again to ensure it remains a strict precondition to reach the target states.
At the next iteration of refinement, \pname proposes to propagate this predicate further inside the branch that parses the \texttt{IDAT} chunk (\redcircle{3}) and creates a new helper variable \texttt{found\_plte}.
That is because the specification of the PNG file requires that the optional \texttt{PLTE} chunk must appear before the \texttt{IDAT} chunk.
Therefore, by the time the parsing procedure reaches the \texttt{IDAT} chunk, we can already determine whether the input PNG file contains a \texttt{PLTE} chunk.

Such a finalized predicate benefits all fuzzers, as it can prioritize inputs with PLTE properties early in the execution and save the fuzzers from considering PLTE-irrelevant PNG images.
With such a predicate, AFLGo \citep{bohme2017aflgo} gained an impressive 8\(\times\) speedup in triggering this vulnerability with only a three-line change in \texttt{png\_read\_info}.

\section{Methodology}
\label{sec:methodology}

\begin{table*}[!t]
\caption{Toolset for synthesizer agent}
\label{tab:apis}
\renewcommand{\arraystretch}{1.1}
\small
\centering
\begin{tabular}{llll}
\toprule
    \textbf{API} & \textbf{Description} & \textbf{Output Example} & \textbf{Stats}\\
\midrule
\textbf{Search} \\
\texttt{class(cls, [file])} & Search for class or structure \texttt{cls} in the codebase or \texttt{file} & \texttt{typedef struct png\_XYZ \{ ... \}} & 13\%\\
\texttt{method(m, [file])} & Search for method or function \texttt{m} in the codebase or \texttt{file} & \texttt{void png\_read(png\_ptr, ...) \{...\}} & 17\%\\
\texttt{symbol(s, [file])} & Search for symbol in the codebase or \texttt{file} & \texttt{TIFFDataType dtype = TIFF\_BYTE;} & 10\%\\
\texttt{code(c, [file])} & Search for code snippet \texttt{c} in the codebase or \texttt{file} & \texttt{raw2tiff.c:302:memcpy(buf, ...) } & 21\%\\
\midrule
\textbf{Graph} \\
\texttt{callers(f)} & Get the caller names of function \texttt{f} in the codebase & [exif\_iif\_add\_tag, exif\_iif\_add\_fmt, ...] & 13\% \\
\texttt{callees(f)} & Get the callee names of function \texttt{f} in the codebase & \texttt{[php\_strnlen, estrndup, php\_ifd\_double,...]} & 6\%\\
\texttt{references(s)} & Get all references of the symbol \texttt{s} in the codebase & \texttt{[misc.c:2670:if(sig\_num!=SIGALRM)\{, ...]} & 8\%\\
\midrule
\textbf{Listing} \\
\texttt{files(dir)} & List all file names under the given path \texttt{dir} & \texttt{[caf.c, chunk.c, common.h, ...]} & 11\%\\
\texttt{classes(file)} & list all classes and structures defined in \texttt{file} & \texttt{[SF\_INFO, sf\_private\_tag, SF\_VIRTUAL\_IO, ...]} & 5\%\\
\texttt{methods(file)} & List all methods and functions defined in \texttt{file} & \texttt{[exif\_get\_tag\_ht, ifd\_get32s, ptr\_offset, ...]} & 6\%\\
\bottomrule
\end{tabular}
\end{table*}

\Cref{fig:overview} illustrates the high-level workflow of \pname.
Given a specified canary \(\psi\) and the program codebase \(P\), \pname outputs a new codebase \(P^{\prime}\) instrumented by a set of predicates \(\Phi \), where each \(\phi\in \Phi\) is represented by a branch condition with early exit if the predicate condition is not satisfied.
The synthesizer is responsible for generating candidate predicates $\Phi$, while the validator ensures that $\Phi$ are both syntactically valid and semantically consistent with the $\psi$, \ie relaxing the canary conditions.
The fuzzer will run on $P'$ to receive more progress feedback to the canary while enjoying the early termination.
In the following, we formalize the task and then elaborate on each design component in \pname.

\subsection{Formalization}
\label{subsec:methodology-terminology}

We use the notation $P \Downarrow_x S$ to indicate that the program $P$, when executed with input \(x\in X\) sampled from $P$'s input space $X$, can reach a set of program states \(S\).
Assume triggering a vulnerability \(v\) can be characterized as reaching a set of program states \(S_v\).
Given the program under fuzz ($P'$) instrumented by $\Phi$, we need to make sure that the early termination introduced in $\Phi$ preserves the same fuzzing behavior on $P'$ as that of $P$, \ie $\Phi$ do not reject any $x\in X$ that would have reached $S_v$ in $P$.
To this end, we formally define \textit{fuzzing admissibility}.

\begin{definition}
For a vulnerability \(v\), the program \(P^{\prime}\) is fuzzing admissible to \(P\), iff \( \forall x\in X, P^{\prime} \Downarrow_x S_v \implies P \Downarrow_x S_v\).
\end{definition}

While it is unlikely to ensure $P'$ is fuzzing admissible to $P$ in general without a pre-defined target vulnerability $v$, we show such a property is well-defined when $v$ is given and explicitly represented by the canary $\psi$.

A program predicate \(\phi: s \rightarrow \{\text{True}, \text{False}\}\) is a boolean mapping over the program state space.
Concretely, any conditions inside the branch statements, \eg \texttt{if}, \texttt{while}, or \texttt{assert}, can be regarded as predicates, as they evaluate a Boolean expression over the program state.
We define a special class of predicates, namely \textit{canaries}, to characterize vulnerable states:

\begin{definition}
A vulnerability canary \(\psi\) is a predicate s.t. \(\forall s \in S_v \Leftrightarrow \psi(s) = \text{True}\).
\end{definition}

The goal of directed fuzzing towards $S_v$ is equivalent to finding inputs that satisfy the canary $\psi$.
To provide semantically meaningful guides to directed fuzzing, we may instrument the program $P$ with an additional predicate \(\phi\).
Such instrumentation is admissible if and only if $\phi$ is a \textit{relaxation} of the true vulnerability canary $\psi$:

\begin{definition}\label{def:relaxation}
A predicate \(\phi\) is the relaxation of canary \(\psi\), if $\forall s \in S_v, \psi(s) = \text{True} 
\implies 
\phi(s) = \text{True}$.
\end{definition}

By definition, $P'$ instrumented by \(\Phi\) is fuzzing admissible if every predicate $\phi$ is a relaxation of $\psi$.
This suggests that fuzzing $P'$ is equivalent to fuzzing $P$ while enjoying the additional guidance and early termination introduced by the instrumented predicates.

\begin{theorem}
The instrumented program \(P'\) is fuzzing admissible to \(P\), if $P'$ is instrumented with \(\Phi\), where every \(\phi \in \Phi\) is the relaxation of \(\psi\).
\label{thm:fuzzing-admissible}
\end{theorem}

We next illustrate how our agentic synthesizer-validator workflow can produce an \textit{admissible} instrumented program $P'$ that provides \textit{rich semantic feedback} to the directed fuzzer to reach vulnerable states $S_v$.

\subsection{Agent Toolset}
\label{subsec:agent-toolset}

Existing agentic practices for software analyses, such as bug fixing or fault localization, are often equipped with lightweight command-line tools to perform local reasoning\cite{aider,zhang2024autocoderover, yang2024swe-agent}.
This is effective because the root cause of a bug is often spatially close to the observable failure, allowing the agent to retrieve relevant context from a narrow portion of the entire codebase.
However, since our task formulation permits predicates to be synthesized at arbitrary program points, purely lexical tools cannot capture the necessary semantic relationships, and the agent therefore needs additional tools to understand program behaviors.

To equip the agent with such code reasoning capability at arbitrary program points, we provide a suite of tools that \pname can invoke to navigate the entire codebase and retrieve relevant code snippets.
Particularly, in addition to common tools like code search and file listing, \pname integrates specialized tools for traversing program graphs, including call graphs and reference graphs.
Through the call graph APIs, the synthesizer agent can retrieve function call relationships and reason the interprocedural control flow, while the reference graph API allows it to identify variable usages, pointer dereferences, and data access patterns across the codebase.
\Cref{tab:apis} shows the complete toolset and the ratio of their actual usage in our experiments.
Among them, graph traversal accounts for over a quarter of all API invocations.

It is important to note that these graphs are only provided as supplementary references to \pname, as they are all derived statically and may miss program behaviors such as dynamic dispatches and indirect calls.
We observe that the LLM used in \pname is capable of leveraging its learned knowledge to bridge this gap.
For example, in the case of vulnerability TIF002 (see \Cref{tab:main}), \pname successfully resolved the indirect function pointer \texttt{tif->tif\_decoderow} to its concrete implementation \texttt{PixarLogDecode}.

\begin{algorithm}[!t]

\caption{Scaffold of \pname's agentic workflow}
\label{alg:workflow}

\begin{algorithmic}[1]
\Require original program \(P\), vulnerability canary \(\psi\)
\Ensure a target-conditional equivalent program \(P^\prime\)
\State \(\mathcal{C} \gets \textsc{CanaryReasoning}(P, \psi)\) \Comment{list of reasonings}
\State \(\Phi \gets \emptyset\)
\ForAll{\(c \in \mathcal{C}\)}
  \State \(l \gets \textsc{Localize}(c, P)\) \Comment{find the initial program point}
  \State \(n \gets 0\)
  \Repeat
    \State \(\phi_l \gets \textsc{Generate}(l,c,P)\)
    \While{\(\neg \textsc{validate}(\phi_l, \psi, P)\)} \Comment{syntax and semantic}
      \State \(\phi_l \gets \textsc{Generate}(l,\phi,P)\)
    \EndWhile
    \State \(l \gets \textsc{Localize}(\phi, l,c,P)\) \Comment{refine, find a better location}
    \State \(n \gets n + 1\)
  \Until{\(l = \text{None} \lor n > \textsc{MaxIterations}\)}
  \State \(\Phi \gets \Phi \cup \{\phi_l\}\)
\EndFor
\State \(P^\prime \gets \textsc{Instrument}(P, \Phi)\) \Comment{fuzzing admissible program}

\end{algorithmic}
\end{algorithm}

\subsection{Synthesis}
\label{subsec:synthesis}

\Cref{alg:workflow} elaborates on the workflow of \pname.
It iteratively localizes candidate program points and generates the predicates (lines 4–13), then refines and validates them for both syntactic and semantic correctness (line 8). 
Once validated, the predicates are used to instrument the original program so it remains fuzzing-admissible (line 16).

As we demonstrate in \Cref{thm:fuzzing-admissible}, a predicate \(\phi\) should be instrumented at the execution path that can reach the canary \(\psi\).
To synthesize \(\phi\), LLMs must reason about the root cause of the vulnerability and approximate potential execution traces of the program that can trigger it.
The execution trace of a program is often long and complex, involving multiple functions and files.
Therefore, a naive one-shot synthesis approach may not be sufficient, as the synthesizer only retrieves limited context and proposes primitive predicates, \eg an initial predicate generated by \pname (\redcircle{2}~in \Cref{fig:relevant-code}).
To address this, \pname employs an iterative localization-generation refinement workflow.
In each iteration, the synthesizer generates a predicate that preserves the same semantic meaning while moving its placement closer to the program entry.

\pname first synthesizes an initial set of predicates by analyzing the canary and approximates a list of semantic characteristics of the inputs that are likely related to this canary, \eg data structures, types, and properties.
The synthesizer is then prompted to consider these constraints from multiple dimensions.
For example, in \Cref{fig:motivating-example}, besides the predicate, the synthesizer generates multiple constraints, such as requiring that the input PNG file must contain a valid signature.
These approximations are progressively concretized and refined as the synthesizer retrieves more relevant code and reasons about the program execution.

For each approximated characteristic, the synthesizer needs to identify an appropriate program point $l$ where the predicate can be expressed in terms of the variables and expressions in scope.
However, constraining LLM to directly identify the exact program point is often too challenging and unreliable, so the initial stage only asks the synthesizer to select a candidate function rather than a precise program point.
Given the reasoning context generated by the LLM so far, the synthesizer attempts to generate an initial predicate \(\phi_l\) in the candidate function.
Once \(\phi_l\) passes the validation stage (\Cref{subsec:validation}), we refine it in the following iterations.
The goal of such a refinement is to move the predicate to a closer program point to the program entry, while preserving the same semantic meaning as the original predicate, \eg checking for the same characteristics.
This allows the program to reject invalid inputs sooner in the execution, enabling the fuzzer to explore more valid mutations within the same time budget.

It is worth noting that with the refinement iteration, a predicate can be all the way refined toward the fuzzing harness.
In some cases, \pname can indeed effectively synthesize predicates at the program entry, making it similar in spirit to automated harness generation works such as HGFuzzer~\cite{xu2025hgfuzzer} and InputBlaster~\cite{liu2024inputblaster}.
However, in most cases, the input constraints cannot be directly accessed at the harness level.
For example, in \Cref{subsec:motivating-example}, verifying the presence and content of a \texttt{PLTE} chunk requires parsing internal structures of the input that are only accessible deeper in the program.
This necessitates placing predicates at intermediate program points.

\subsection{Validation}
\label{subsec:validation}

The validation step is critical to preserving the correctness of the instrumented program and ensuring that the inserted predicate $\phi_l$ maintains the instrumented program's fuzzing admissibility. 
As shown in \Cref{alg:workflow} lines 7 to 9, the validator takes as input the candidate predicate $\phi$, the vulnerability canary $\psi$, and the program $P$, and validates whether the predicate is both syntactically and semantically correct.
If the predicate passes both checks, \pname will refine the predicate by exploring potentially better program points closer to the program entries (\Cref{subsec:synthesis}).
If it fails, \pname will self-reflect and attempt to regenerate the predicate, using diagnostic feedback collected from the validator.

\para{Syntax validation}
The first component of validation is syntactic checking.
A predicate that fails to compile cannot be used in a fuzzing campaign, regardless of its intended semantics.
To verify this, the program is instrumented by the predicate \(\phi\) at the designated program point $l$ and invokes the project’s build system using a predefined command.
If the build fails, the associated compiler error messages will be sent to the synthesizer to repair.
This diagnostic information typically involves undeclared symbols, type mismatches, or malformed expressions.

\para{Semantic validation}
The second component is semantic validation, which confirms that the predicate $\phi$ strictly relaxes $\psi$.
It ensures that the predicate does not reject any execution paths that can reach the vulnerability (Theorem~\ref{thm:fuzzing-admissible}).
As we demonstrated in \Cref{def:relaxation}, such validation requires enumerating all possible program inputs, which cannot be done within an acceptable time budget.
Therefore, we utilize symbolic execution to find counterexamples of the relaxation.

\begin{theorem}\label{thm:relaxation-validation}
A predicate \(\phi\) is not a relaxation of \(\phi'\), if there exists \( P\Downarrow_x s\), s.t. \( \phi(s) = \text{False} \land \phi'(s) = \text{True}\)
\end{theorem}

Specifically, the predicate is not a strict relaxation of the canary if the symbolic execution can find a path that satisfies the negated predicate \(\neg \phi\) while the canary \(\psi\) still evaluates to true.

It is natural to employ symbolic execution as a formal checker for \Cref{thm:relaxation-validation} to ensure fuzzing admissibility.
However, it often incurs substantial overhead by exploring irrelevant execution paths\cite{cadar2008klee}, \eg branches that are unrelated to either the synthesized predicate or the target canary.
This excessive path exploration can lead to prohibitively long validation time, incurring additional overhead of deploying \pname.
To mitigate this inefficiency, we adopt a strategy inspired by Chopper\cite{trabish2018chopper} to skip unrelated paths and target only the exploration of paths according to our selection.
Specifically, we perform a lightweight reachability analysis on the CFG and prune nodes that are not reachable from either the predicate or the canary.
\pname then initiates symbolic execution to explore the path between the \(\neg \phi\) and \(\psi\).
Overall, the semantic validation offers a sound guarantee that the relaxation is harmless, but it remains unable to measure the effectiveness of the generated predicates.

\section{Evaluation}\label{sec:evaluation}
We aim to answer the following research questions:

\begin{table*}[!t]

\centering
\small
\setlength{\tabcolsep}{1.2pt} %
\caption{
TTE for each vulnerability in the Magma benchmark across different fuzzers.
T.O. indicates that the fuzzer cannot find the vulnerability within 24 hours.
\(\varnothing\) indicates that the fuzzer either could not compile the target program or the preprocessing step exceeded 24 hours.
Numbers in \sig{bold} indicate statistical significance ($p \le 0.05$).
}
\label{tab:main}
\begin{tabular}{l @{\hskip 8pt} ll @{\hskip 12pt} ll @{\hskip 12pt} ll @{\hskip 12pt} ll @{\hskip 12pt} ll @{\hskip 12pt} ll @{\hskip 12pt} ll @{\hskip 12pt} ll}
\toprule
\multirow{2}[2]{*}{\textbf{Vul ID}} &
\multicolumn{2}{c}{\hspace{-1.5em}\textbf{AFLGo}} &
\multicolumn{2}{c}{\hspace{-1.5em}\textbf{SelectFuzz}} &
\multicolumn{2}{c}{\hspace{-1.5em}\textbf{Beacon}} &
\multicolumn{2}{c}{\hspace{-1.5em}\textbf{Titan}} &
\multicolumn{2}{c}{\hspace{-1.5em}\textbf{AFL++}} &
\multicolumn{2}{c}{\hspace{-1.5em}\textbf{AFL}} &
\multicolumn{2}{c}{\hspace{-1.5em}\textbf{MOPT}} &
\multicolumn{2}{c}{\textbf{Fox}} \\

\cmidrule[\heavyrulewidth](l{0.5em}r{1.5em}){2-3}
\cmidrule[\heavyrulewidth](l{0.5em}r{1.5em}){4-5}
\cmidrule[\heavyrulewidth](l{0.5em}r{1.5em}){6-7}
\cmidrule[\heavyrulewidth](l{0.5em}r{1.5em}){8-9}
\cmidrule[\heavyrulewidth](l{0.5em}r{1.5em}){10-11}
\cmidrule[\heavyrulewidth](l{0.5em}r{1.5em}){12-13}
\cmidrule[\heavyrulewidth](l{0.5em}r{1.5em}){14-15}
\cmidrule[\heavyrulewidth](lr){16-17}
& \textbf{Origin} & \textbf{\pname} & \textbf{Origin} & \textbf{\pname} & \textbf{Origin} & \textbf{\pname} & \textbf{Origin} & \textbf{\pname} & \textbf{Origin} & \textbf{\pname} & \textbf{Origin} & \textbf{\pname} & \textbf{Origin} & \textbf{\pname} & \textbf{Origin} & \textbf{\pname} \\
\midrule
PNG003 & 4 & \insig{4} & 3 & \insig{3} & 5 & \insig{5} & 8 & \insig{8} & 12 & \gc{15}\sig{8} & 3 & \insig{3} & 4 & \insig{4} & 5 & \insig{5} \\
PNG006 & T.O. & T.O. & T.O. & T.O. & T.O. & T.O. & T.O. & T.O. & 106 & \gc{15}\insig{63} & T.O. & T.O. & T.O. & T.O. & 6978 & \gc{25}\sig{1582} \\
PNG007 & 72536 & \gc{30}\sig{9052} & 58770 & \gc{30}\sig{8537} & 7764 & \gc{25}\sig{1659} & 17561 & \gc{15}\sig{11954} & 53104 & \gc{8}\sig{41101} & 55351 & \gc{25}\sig{18358} & 42811 & \gc{8}\insig{37522} & 4651 & \gc{8}\sig{4009} \\
SND001 & T.O. & \gc{70}\sig{419} & 7764 & \gc{70}\sig{5} & 432 & \gc{50}\sig{13} & 115 & \gc{40}\sig{8} & 451 & \gc{50}\sig{19} & 75940 & \gc{70}\sig{338} & 285 & \gc{50}\sig{12} & 447 & \gc{40}\sig{27} \\
SND005 & 1015 & \gc{30}\sig{205} & 102 & \gc{50}\sig{4} & 95 & \gc{30}\sig{16} & 30412 & \gc{50}\sig{1046} & 1233 & \gc{30}\sig{193} & 796 & \gc{60}\sig{9} & 53 & \gc{30}\sig{9} & 82 & \gc{30}\sig{15} \\
SND006 & T.O. & \gc{50}\sig{3899} & 8519 & \gc{70}\sig{12} & T.O. & T.O. & 26171 & \gc{50}\sig{905} & 7026 & \gc{70}\sig{24} & T.O. & \gc{50}\sig{2251} & 328 & \gc{40}\sig{17} & 15422 & \gc{50}\sig{717} \\
SND007 & T.O. & \gc{50}\sig{3468} & 7785 & \gc{70}\sig{62} & T.O. & T.O. & 310 & \gc{40}\sig{28} & 810 & \gc{50}\sig{34} & T.O. & \gc{50}\sig{3163} & 360 & \gc{50}\sig{16} & 3226 & \gc{50}\sig{129} \\
SND017 & 4453 & \gc{25}\sig{1021} & 3235 & \gc{20}\sig{1518} & 4861 & \gc{60}\sig{58} & 40955 & \gc{70}\sig{256} & 841 & \gc{25}\sig{196} & 3162 & \gc{25}\sig{833} & 33 & \gc{15}\sig{20} & 2117 & \gc{40}\sig{178} \\
SND020 & 3805 & \gc{20}\sig{1291} & 694 & \gc{8}\sig{542} & 5683 & \gc{70}\sig{41} & 70432 & \gc{70}\sig{604} & 640 & \insig{204} & 2916 & \gc{20}\sig{1266} & 326 & \gc{25}\sig{103} & 2068 & \gc{40}\sig{176} \\
SND024 & T.O. & \gc{50}\sig{2577} & 5957 & \gc{70}\sig{18} & T.O. & T.O. & 1074 & \gc{70}\sig{8} & 361 & \gc{40}\sig{18} & T.O. & \gc{50}\sig{2391} & 326 & \gc{50}\sig{14} & 2642 & \gc{60}\sig{32} \\
TIF002 & T.O. & T.O. & T.O. & T.O. & T.O. & T.O. & T.O. & T.O. & 72896 & \insig{69963} & T.O. & T.O. & T.O. & T.O. & T.O. & T.O. \\
TIF005 & T.O. & T.O. & T.O. & T.O. & T.O. & T.O. & T.O. & T.O. & 420 & \gc{15}\sig{260} & T.O. & T.O. & T.O. & T.O. & 4744 & \gc{25}\sig{990} \\
TIF006 & 80998 & \gc{25}\sig{17774} & T.O. & T.O. & 35493 & \gc{30}\sig{6715} & 66692 & \gc{8}\sig{65412} & 970 & \gc{20}\sig{336} & 52926 & \gc{20}\sig{25366} & 39810 & \insig{45124} & 22380 & \gc{20}\sig{7607} \\
TIF007 & 18513 & \gc{30}\sig{2107} & 1410 & \gc{25}\sig{413} & 220 & \gc{20}\sig{76} & 146 & \gc{8}\sig{109} & 50 & \gc{20}\sig{26} & 15022 & \gc{40}\sig{1172} & 155 & \gc{20}\sig{56} & 8868 & \gc{50}\sig{290} \\
TIF009 & 32556 & \gc{15}\sig{16767} & 12067 & \gc{20}\sig{4144} & T.O. & T.O. & T.O. & T.O. & 17483 & \gc{25}\sig{5234} & 19302 & \gc{8}\sig{14098} & 26049 & \insig{15622} & 7696 & \gc{20}\sig{3406} \\
TIF012 & 47424 & \gc{20}\sig{18689} & 9118 & \gc{15}\sig{4787} & 18782 & \gc{15}\sig{12274} & 2922 & \gc{8}\sig{2620} & 1731 & \gc{15}\sig{1122} & 28884 & \insig{16318} & 3639 & \gc{25}\sig{953} & 1257 & \gc{20}\sig{637} \\
TIF014 & 66000 & \gc{30}\sig{8134} & 58851 & \gc{30}\sig{7833} & 19035 & \gc{20}\sig{9672} & 44384 & \gc{25}\sig{14360} & 2555 & \gc{25}\sig{682} & 66072 & \gc{30}\sig{7303} & 1660 & \gc{20}\sig{788} & 4666 & \gc{8}\sig{3275} \\
LUA004 & T.O. & T.O. & T.O. & T.O. & T.O. & T.O. & 69594 & \gc{8}\sig{67442} & 20046 & \gc{15}\sig{11939} & 44954 & \gc{20}\sig{22434} & 12401 & \gc{30}\sig{2098} & 35998 & \gc{30}\sig{6937} \\
XML001 & T.O. & T.O. & T.O. & T.O. & T.O. & T.O. & T.O. & T.O. & 3720 & \gc{20}\sig{1330} & T.O. & T.O. & 37726 & \gc{20}\sig{16801} & T.O. & T.O. \\
XML003 & T.O. & T.O. & T.O. & T.O. & 44056 & \gc{25}\sig{13295} & 53689 & \insig{36365} & 2373 & \gc{15}\sig{1394} & T.O. & T.O. & 56501 & \gc{25}\sig{18636} & 12252 & \gc{15}\sig{6844} \\
XML009 & T.O. & \insig{77513} & T.O. & \gc{70}\sig{56} & 32585 & \gc{25}\sig{6748} & 17942 & \gc{8}\sig{13879} & 2668 & \gc{15}\sig{1452} & T.O. & \insig{22753} & 706 & \insig{670} & 10827 & \gc{8}\sig{8766} \\
XML017 & 17 & \gc{20}\sig{8} & 7 & \gc{15}\sig{9} & 10 & \insig{15} & 23 & \gc{15}\sig{13} & 52 & \gc{25}\sig{15} & 26 & \gc{25}\sig{7} & 6 & \gc{15}\sig{16} & 77 & \gc{20}\sig{35} \\
SSL001 & T.O. & T.O. & 72078 & \gc{8}\sig{62637} & $\varnothing$ & $\varnothing$ & $\varnothing$ & $\varnothing$ & 23533 & \gc{40}\sig{1583} & 81838 & T.O. & 43952 & \gc{20}\sig{16033} & 66778 & \gc{25}\sig{17764} \\
SSL003 & 222 & \gc{8}\sig{231} & 73 & \gc{8}\sig{69} & $\varnothing$ & $\varnothing$ & $\varnothing$ & $\varnothing$ & 193 & \gc{8}\sig{178} & 90 & \gc{8}\sig{63} & 96 & \gc{8}\sig{79} & 225 & \insig{199} \\
PHP004 & 165 & \gc{8}\sig{269} & 11 & \insig{11} & $\varnothing$ & $\varnothing$ & $\varnothing$ & $\varnothing$ & 66156 & \gc{15}\sig{34974} & 82 & \gc{15}\sig{89} & 760 & \gc{25}\sig{185} & $\varnothing$ & $\varnothing$ \\
PHP009 & 99 & \gc{8}\sig{203} & 185 & \gc{8}\sig{153} & $\varnothing$ & $\varnothing$ & $\varnothing$ & $\varnothing$ & 11890 & \gc{20}\sig{5677} & 143 & \insig{179} & 650 & \insig{596} & $\varnothing$ & $\varnothing$ \\
PHP011 & $\varnothing$ & $\varnothing$ & 26 & \gc{8}\sig{23} & $\varnothing$ & $\varnothing$ & $\varnothing$ & $\varnothing$ & 1626 & \insig{934} & 108 & \insig{76} & 19 & \gc{15}\sig{13} & $\varnothing$ & $\varnothing$ \\
SQL018 & 49520 & \gc{15}\sig{30855} & 53397 & \gc{30}\sig{6748} & T.O. & T.O. & 78238 & \gc{15}\sig{42275} & 10355 & \gc{20}\sig{4961} & 71038 & \gc{20}\sig{32073} & 22656 & \insig{28101} & 15498 & \gc{15}\sig{8159} \\
\midrule
\textbf{Speedup} & \multicolumn{2}{c}{\textbf{17.0\(\times\)}} & \multicolumn{2}{c}{\textbf{214.2\(\times\)}} & \multicolumn{2}{c}{\textbf{22.1\(\times\)}} & \multicolumn{2}{c}{\textbf{28.0\(\times\)}} & \multicolumn{2}{c}{\textbf{15.3\(\times\)}} & \multicolumn{2}{c}{\textbf{20.4\(\times\)}} & \multicolumn{2}{c}{\textbf{5.5\(\times\)}} & \multicolumn{2}{c}{\textbf{10.6\(\times\)}}\\
\bottomrule
\end{tabular}
\end{table*}

\begin{enumerate}[label=\textbf{RQ\arabic*},leftmargin=3em]
\item\label{rq:effect} \textbf{Effectiveness}: How effective is \pname in accelerating the generation of PoC inputs for given target vulnerabilities?
\item\label{rq:cost} \textbf{Cost and performance}: What is the time cost and token cost for deploying \pname?
\item\label{rq:ablation} \textbf{Ablations}: How do the individual components of \pname contribute to its overall performance?
\item\label{rq:impact} \textbf{Security impact}: How can \pname assist in real-world vulnerability detection scenarios?
\end{enumerate}

\subsection{Setup}
\label{subsec:setup}

\para{Dataset}
We evaluate \pname on the Magma fuzzing benchmark \cite{hazimeh2020magma}, which includes a diverse set of real-world vulnerabilities selected from nine popular open-source software projects.
We also consider popular libraries in the wild, \eg VLC, to evaluate the capabilities of \pname in finding real-world vulnerabilities (see \Cref{subsec:case_study}).
Each vulnerability in Magma is represented by a canary statement (Section~\ref{subsec:motivating-example}), placed within the target function (\Cref{fig:motivating-example}).
The canary condition is manually curated to specify the state necessary to trigger the vulnerability and logs the time when it is satisfied.
Note that while our evaluation assumes the canary is provided, \pname can be extended to other forms of targets, including vulnerability reports, patches, or static analysis results, by introducing an additional processing step (\Cref{subsec:case_study}).

\para{Baselines}
We select eight fuzzers (See \Cref{tab:fuzzers}) as our baseline, covering both the state-of-the-art directed and coverage-guided ones.
For each fuzzer, we use the latest stable version available at the time of writing.
Note that \pname alone is not a standalone fuzzer, but focuses exclusively on code transformations for the target software.
Therefore, it is agnostic to fuzzer implementations and can complement any fuzzers (\Cref{fig:overview}).

\begin{table}[!t]

\caption{Overview of selected fuzzers in the evaluation}
    \label{tab:fuzzers}

\small
\setlength{\tabcolsep}{4pt}

\centering
\begin{tabular}{lll}
\toprule
\textbf{Fuzzer} & \textbf{Category} & \textbf{Description} \\
\midrule
AFLGo\cite{bohme2017aflgo} & directed & Distance-based seeds scheduling \\
    SelectFuzz\cite{luo2023selectfuzz} & directed & Selective path exploration \\
    Beacon\cite{huang2022beacon} & directed & Fuzzer with efficient path pruning \\
    Titan\cite{huang2024titan} & directed & Targets correlations inference \\
\midrule
    AFL\cite{zalewski2020afl} & coverage-guided & Evolutionary mutation strategies \\
    AFL++\cite{fioraldi2020afl++} & coverage-guided & Community-enhanced AFL \\
    MOPT\cite{lyu2019mopt} & coverage-guided & Fuzzer with Swarm Optimization \\
    Fox\cite{she2024fox} & coverage-guided & Online stochastic control \\
\bottomrule
\end{tabular}
\end{table}

\para{Metrics}
We follow the existing directed fuzzing approaches by adopting the Time To Exposure (TTE) to measure the performance of the baseline fuzzers and the improvement introduced by integrating \pname.
In the Magma benchmark, TTE measures the time taken by a fuzzer to find the input that satisfies the canary condition.
To mitigate the inherent randomness introduced by fuzzing, we follow \citet{hazimeh2020magma} by executing 10 independent fuzzing trials per vulnerability sample and report the average TTE.
We also employ the Mann-Whitney U Test\cite{mann1947test} to demonstrate the statistical significance ($p$-value) of the results.

\para{Implementations}
We run all the fuzzers with the same initial seed inputs provided in Magma to ensure a fair comparison.
Each fuzzing trial is capped at 24 hours, so those that fail to find the triggering input are recorded at this maximum duration.
This potentially underestimates the actual TTE for baseline fuzzers, but it offers a lower-bound estimation.
Therefore, the actual improvement brought by \pname can be even larger.
To eliminate hardware and system-related discrepancies, all experiments are conducted on a dedicated cluster, where each server comes with an Intel Xeon Gold 6126 CPU and 128GB of RAM running Ubuntu 20.04.

We implement \pname using the PydanticAI framework \cite{pydanticai}.
The synthesizer's tooling is primarily built on top of Multiplier \cite{multiplier}.
We leverage the SVF static analysis framework \cite{sui2016svf} to perform a lightweight reachability analysis and use KLEE \cite{cadar2008klee} as the symbolic execution engine to perform semantic validation.
We use \texttt{o3-mini-2025-01-31} as the default model with \texttt{reasoning level} set to \texttt{medium}, but we also evaluate other LLMs in \Cref{subsec:ablation_study}.

\subsection{RQ1: Vulnerabilities Reproduction}
\label{subsec:main-results}

We apply \pname on both directed fuzzers and coverage-guided fuzzers, comparing the performance with and without \pname.
The results are shown in \Cref{tab:main}.

In summary, \pname consistently improves all kinds of fuzzers for vulnerability reproduction.
When integrating \pname to directed fuzzers, AFLGo\cite{bohme2017aflgo}, SelectFuzz\cite{luo2023selectfuzz}, Beacon\cite{huang2022beacon}, and Titan\cite{huang2024titan} achieve 17.0\(\times\), 214.2\(\times\), 22.1\(\times\), and 28.0\(\times\) faster TTE on average, with five more vulnerabilities found for AFLGo and one more vulnerability found for SelectFuzz than those without \pname within the fixed 24-hour time window.

For coverage-guided fuzzers, integrating \pname also yields substantial gains: AFL++\cite{fioraldi2020afl++}, AFL\cite{zalewski2020afl}, MOPT\cite{lyu2019mopt}, and FOX\cite{she2024fox} achieve 15.3\(\times\), 20.4\(\times\), 5.5\(\times\), and 10.6\(\times\) faster TTE on average, with four more vulnerabilities found for AFL than those without \pname within the 24-hour time window.

\subsection{RQ2: Cost Analysis}
\label{subsec:cost-analysis}
We measure the time cost introduced by \pname and the token cost incurred by LLM inference.

\para{Time cost}
In \Cref{subsec:main-results}, we focus on measuring TTE, \ie the time required to trigger the target canary during fuzzing.
However, evaluating only the fuzzing phase can be misleading when assessing the overall effectiveness of a directed fuzzer.
Many directed fuzzers perform expensive static analysis on the target program before fuzzing begins\cite{bohme2017aflgo, huang2024titan, huang2024halo}.
Likewise, \pname requires additional preprocessing steps, including codebase indexing, symbolic validation of predicates, and agentic predicate synthesis.
Although such preprocessing costs are often treated as a one-time effort amortized over fuzzing, we find that some approaches incur excessive analysis time, sometimes even longer than the time required to discover the bug by the fuzzer.

We report the detailed preprocessing time overhead incurred by \pname and baseline directed fuzzers.
As shown in \Cref{tab:cost}, the overhead of the baselines exponentially grows with the size of the codebase.
In contrast, \pname relies only on lightweight analysis tools, \eg code retrieval, graph traversal, etc., so it remains efficient regardless of the project size.
Particularly, \pname outperforms the best baseline fuzzer SelectFuzz by 4.5\(\times\) when evaluated on the largest program in Magma (PHP).

\para{LLM token cost}
We analyze the LLM token usage to assess the financial feasibility of deploying \pname.
\Cref{tab:cost} shows the token cost for our \pname workflow.
\pname takes 457k tokens (equivalent to \$0.72 USD) to generate predicates for one sample in the Magma benchmark on average.
The monetary token costs show the potential for \pname to act as an affordable step in assisting fuzzing.

\begin{table}[!t]
\setlength{\tabcolsep}{3pt}
\renewcommand{\arraystretch}{1.1}
\centering
\caption{Average deploy cost for \pname and directed fuzzers.
All times are calculated in seconds. T.O. indicates that the fuzzer failed to instrument the target library within 24 hours.
}
\label{tab:cost}
\small
\begin{tabular}{lllllllll}
\toprule
\textbf{Target} & \textbf{PNG} & \textbf{SND} & \textbf{TIF} & \textbf{LUA} & \textbf{XML} & \textbf{SSL} & \textbf{PHP} & \textbf{SQL} \\ \midrule
Size (LoC) & 95k & 83k & 95k & 21k & 320k & 630k & 1.6M & 387k \\
\midrule
Index & 11 & 34 & 82 & 9 & 76 & 146 & 244 & 137 \\
Synthesis & 373 & 331 & 212 & 178 & 215 & 384 & 412 & 349 \\
Validation & 261 & 133 & 231 & 280 & 475 & 824 & 353 & 407 \\
Total & 645 & 498 & 525 & 467 & 766 & 1354 & 1009 & 893 \\
\midrule
\#Tokens (k) & 309 & 303 & 256 & 176 & 653 & 598 & 894 & 467\\
\midrule
AFLGo & 122 & 673 & 2689 & 85 & 5608 & 24799 & T.O. & 15630 \\
SelectFuzz & 84 & 199 & 1167 & 44 & 807 & 2597 & 4554 & 383 \\
Beacon & 64 & 113 & 171 & 35 & 1656 & T.O. & T.O. & 3721 \\
Titan & 96 & 186 & 967 & 49 & 2936 & T.O. & T.O. & 4965 \\
\bottomrule
\end{tabular}
\end{table}

\subsection{RQ3: Ablations}
\label{subsec:ablation_study}

We ablate the design choice of different modules in \pname and study how they generalize to various LLMs in \Cref{tab:ablation}.
We pick five representative vulnerabilities that fall under different vulnerability categories in Common Weakness Enumerations (CWEs).
The evaluation compares two representative mainstream fuzzing tools, AFL++ from the coverage-guided fuzzers and SelectFuzz from the directed fuzzers.

\begin{table}[!t]

\caption{TTEs by ablating different designs and models}
\label{tab:ablation}
\setlength{\tabcolsep}{3pt}
\centering
\small
\begin{tabular}{llllll}
\toprule
& \textbf{PNG007} & \textbf{SND001} & \textbf{TIF012} & \textbf{TIF014} & \textbf{SQL018} \\
\midrule
\multicolumn{6}{l}{Ablate different designs} \\
\midrule
\textbf{AFL++}       &        &        &        &        &        \\
Origin      & 53104  & 451    & 1731   & 2555   & 10355  \\
+Base     & 54830  & 389    & 1904   & 1226   & 11573  \\
+Refine  & T.O.   & 23     & 4417   & 33931  & 54268  \\
+Valid  & \textbf{41101}  & \textbf{19}     & \textbf{1122}   & \textbf{682}    & \textbf{4961}   \\
\cmidrule(lr){1-6}
\textbf{SelectFuzz}  &        &        &        &        &        \\
Origin        & 58770  & 7764   & 9118   & 58851  & 53397  \\
+Base        & 46207  & 2640   & 8938   & 12630  & 48059  \\
+Refine  & T.O.   & 6      & 10805  & T.O.   & T.O.   \\
+Valid  & \textbf{8537}   & \textbf{5}      & \textbf{4787}   & \textbf{7833}   & \textbf{6748}   \\
\midrule
\multicolumn{6}{l}{Ablate different LLMs} \\
\midrule
\textbf{AFL++}       &        &        &        &        &        \\
Origin        & 53104  & 451    & 1731   & 2555   & 10355  \\
with \texttt{o3-mini}     & 41101  & \textbf{19}     & 1122   & \textbf{682}    & \textbf{4961}   \\
with \texttt{Deepseek R1} & 45104  & 53 & \textbf{992}    & 1439   & 5433   \\
with \texttt{Gemini Flash 2.0}   & \textbf{38215}  & 130    & 1517   & 2454   & 5274   \\
\cmidrule(lr){1-6}
\textbf{SelectFuzz}  &        &        &        &        &        \\
Origin        & 72078  & 7764   & 9118   & 58851  & 53397  \\
with \texttt{o3-mini}    & 8537   & \textbf{5}      & 4787   & 7833   & 6748   \\
with \texttt{Deepseek R1} & 23699  & 6      & \textbf{3562}   & \textbf{3518}   & 7023   \\
with \texttt{Gemini Flash 2.0}    & \textbf{7020}   & 12     & 7206   & 40332  & \textbf{3921} \\
\bottomrule
\end{tabular}
\end{table}

\para{Effectiveness of each design component}
We ablate the individual design and measure the average TTE obtained by the resulting predicates.
We begin with the \emph{Base} setting, where we only keep the initial synthesized predicate (\Cref{alg:workflow} line 7).
Next, we apply the refinement strategy, allowing the synthesized predicate to propagate to a better program point without validating its semantic correctness.
Finally, we evaluate the full pipeline of \pname including both the syntax and semantic validation.

\Cref{tab:ablation} shows that the design choices in \pname consistently improve the performance.
Specifically, we observe that while predicates generated under the base setting may occasionally accelerate fuzzing, they are not consistently beneficial.
Without validation, \pname sometimes generates false predicates to the program, as evidenced by the cases PNG007, TIF014, and SQL018.

\para{Varying models}
Besides the default \texttt{o3-mini} model, we consider \texttt{Deepseek R1}\cite{deepseek-ai2025deepseek-r1} and \texttt{Gemini 2.0 Flash}\cite{gemini2flash} to study the generality of our agentic framework to different LLM architectures.
\Cref{tab:ablation} shows that \pname generalizes to different LLM architectures, except for one case where Gemini fails to bring significant improvement to TIF014.
We investigate this case and find that Gemini failed to elevate the generated predicate to the caller closer to the input, such that the additional overhead introduced by evaluating the predicates outweighs the benefit it brings to the fuzzing.

\subsection{RQ4: Detecting New Vulnerabilities}
\label{subsec:new-vuls}
We integrate \pname into a real-world vulnerability detection workflow and apply the pipeline to a set of well-fuzzed targets, such as \texttt{VLC}\cite{videolan_vlc}, \texttt{libming}\cite{libming}, \texttt{libarchive}\cite{libarchive}, and \texttt{tcpreplay} \cite{tcpreplay}. 
To construct meaningful fuzzing targets, we generate new canaries through two primary approaches.
First, we leverage alerts produced by the static analysis tool SVF\cite{sui2016svf}, which identify potentially vulnerable program points based on memory access patterns, aliasing behavior, or use-after-free risks.
Second, we derive reachability canaries for program points associated with previously patched bugs.
The rationale behind this strategy is that many real-world bugs occur in clusters or evolve from incomplete fixes\cite{wang2020not}.

We launch 48 fuzzing samples in total, with each fuzzing campaign lasting for 24 hours.
We found nine previously undiscovered bugs, including memory leaks, use-after-free, null pointer dereference, and out-of-bound memory access (\Cref{tab:new-vul}). 
At the time of writing, we responsibly reported all the new vulnerabilities to the relevant maintainers, and three of the bugs already have drafted fixes.
To further evaluate the efficiency improvement, we also repeat the fuzzing campaign by fuzzers without \pname.
The results show that without \pname, AFL++ alone can only detect 2/9 vulnerabilities, and SelectFuzz alone can only detect 5/9 vulnerabilities.
\Cref{subsec:case_study} elaborates on one newly detected vulnerability.

\begin{table}[!t]
\small
\centering
\setlength{\tabcolsep}{4pt}
\caption{New vulnerabilities detected by fuzzers with \pname.
For each newly found vulnerability, we run AFL++ and SelectFuzz with and without \pname.
}
\label{tab:new-vul}

\begin{tabular}{llllll}
\toprule
\multirow{2}[2]{*}{\textbf{Bug ID}} & \multirow{2}[2]{*}{\textbf{Type}} & \multicolumn{2}{c}{\textbf{AFL++}} & \multicolumn{2}{c}{\textbf{SelectFuzz}} \\
\cmidrule[\heavyrulewidth](lr){3-4}
\cmidrule[\heavyrulewidth](lr){5-6}
& & \textbf{origin} & \textbf{\pname} & \textbf{origin} & \textbf{\pname} \\
\midrule
VLC-29163 & Memory leak & T.O.& 30847 & T.O. & 26317  \\
VLC-29162 & OOB access & T.O. & 74835 & T.O. & T.O. \\
VLC-29238 & Memory leak & T.O. & 21085 & 53872 & 24766\\
VLC-29239 & Use-after-free & 43680 & 3946 & 22983 & 5405 \\
libming-365 & Null deref & T.O. & 80241 & T.O. & T.O. \\
libarchive-hvqg & Null deref & 83622 & 34327 & 62748 & 18309 \\
libarchive-fm54 & OOB access & T.O. & 16397 & 46577 & 23280 \\
tcpreplay-pmgq & NULL deref & T.O. & 14219 & T.O. & T.O. \\
tcpreplay-pg55 & OOB access & T.O. & 31526 & 20392 & 21305\\
\bottomrule
\end{tabular}
\end{table}

\subsection{Case Study}
\label{subsec:case_study}

\para{Timeout cases}
\Cref{subsec:main-results} demonstrates that the synthesized predicates can occasionally yield negligible improvement.
For example, the predicates associated with vulnerability \texttt{TIF009} significantly improve the performance of all evaluated fuzzers except for Beacon and Titan.
On average, these predicates bring 2$\times$ speedup to all other fuzzers, while Beacon and Titan always timeout with or without \pname synthesized predicates.

Upon further investigation, we find that their employed termination mechanism can sometimes conflict with the canaries.
Specifically, Beacon and Titan perform static analysis to identify functions potentially reachable from the vulnerability canary, and primarily prioritize program inputs that can reach the statically identified functions.
The predicates synthesized by \pname thus offer only limited guidance when they are put in the functions that are not statically identified as close to the canary.

\para{Canary generation by \pname}
\Cref{subsec:new-vuls} discusses our attempts to generate canaries for arbitrary target states without relying on existing ones.
This shows that \pname need not rely on a pre-defined canary to be applicable.
Instead, it can be extended to automatically generate canaries based on diverse representations of the target states, such as patch and bug descriptions.

\begin{figure}[!t]
    \centering
    \includegraphics[width=\linewidth]{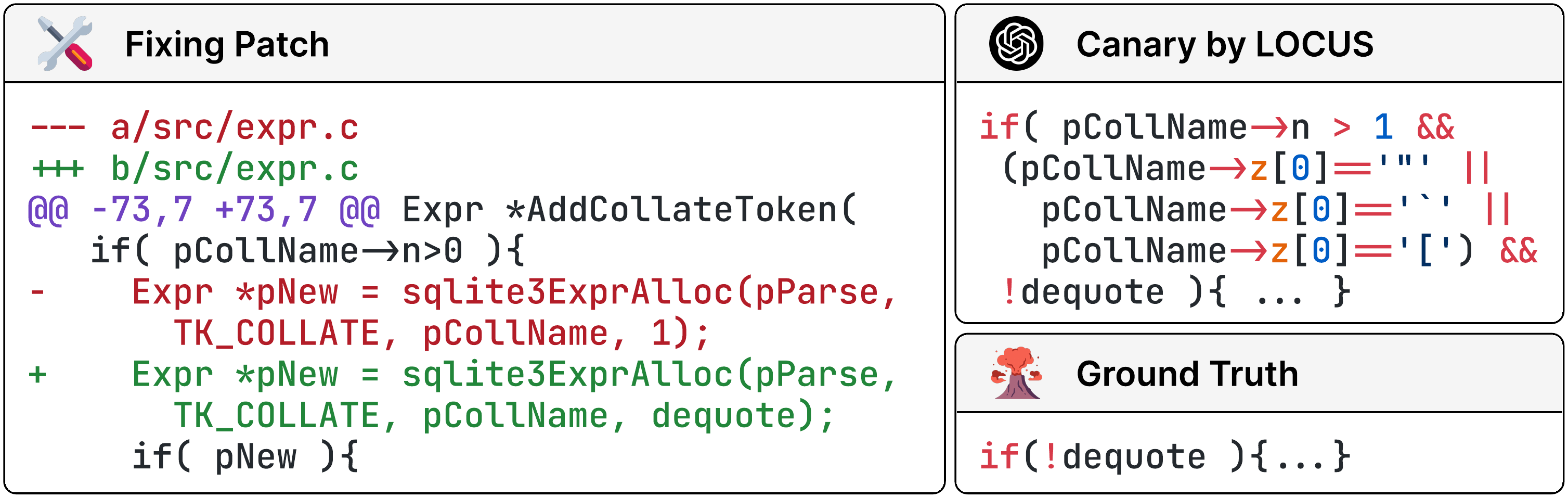}
    \caption{
\pname generates a more precise canary using only the security patch. 
Previously, an incorrectly set dequoting flag enabled access to uninitialized memory.
}
    \label{fig:canary}
\end{figure}

To apply \pname in such settings, we also consider letting \pname generate canary conditions based on available security patches.
Specifically, we iterate over all security patches in Magma and prompt \pname to automate the generation of a canary condition (Magma relies on manual analysis), such that the unpatched version satisfies the generated canary, while the patched version violates it.
By manually checking all the canaries generated by the LLMs, we confirm that 27 out of 28 security patches can be correctly translated into the same condition.

We find the remaining case to be intriguing: the LLM produces a more precise vulnerability canary than the original one in the Magma benchmark, as shown in \Cref{fig:canary}).
Specifically, this vulnerability is found in \texttt{SQLite3} and classified as CWE-908, \ie use of the uninitialized resource. 
When building a collate token, a constant value \texttt{1} is incorrectly passed in instead of the \texttt{dequote} flag.
This allows attackers to access uninitialized memory via forced dequoting, which would lead to memory corruption.
The original vulnerability canary only checks whether dequoting is disabled, and invalidates cases for unquoted names since they would not be dequoted even if the \texttt{dequote} flag were set.
Instead, the canary generated by \pname targets mishandled quoted strings, precisely capturing the underlying vulnerability and potentially reducing false positives from imprecise ground-truth canaries.

\para{Real-world vulnerability}
In this section, we elaborate on the vulnerability \texttt{libarchive-fm54} in \Cref{tab:new-vul}. 
An out-of-bounds error can occur when the program processes a specially crafted RAR file.
This vulnerability stems from insufficient bound checking in the delta filter logic used during decompression.
In the RAR format, filters are lightweight data transformation routines applied to improve compression efficiency.
One common example is the delta filter, which is often used on binary data, such as audio or image streams, to convert absolute values into relative differences, thus making patterns more compressible.
During decompression, the filter operates on two memory regions: the source buffer \texttt{src}, which holds the raw decompressed data, and the destination buffer \texttt{dst}, which stores the filtered output.
This vulnerability will be triggered when \texttt{src} inadvertently points into the \texttt{dst} buffer.
This overlap can lead to undefined behavior, including memory corruption, out-of-bounds access, or program crashes, as the filter may read from or write to memory locations outside the valid range.

\begin{figure}[!t]
\centering
\includegraphics[width=0.9\linewidth]{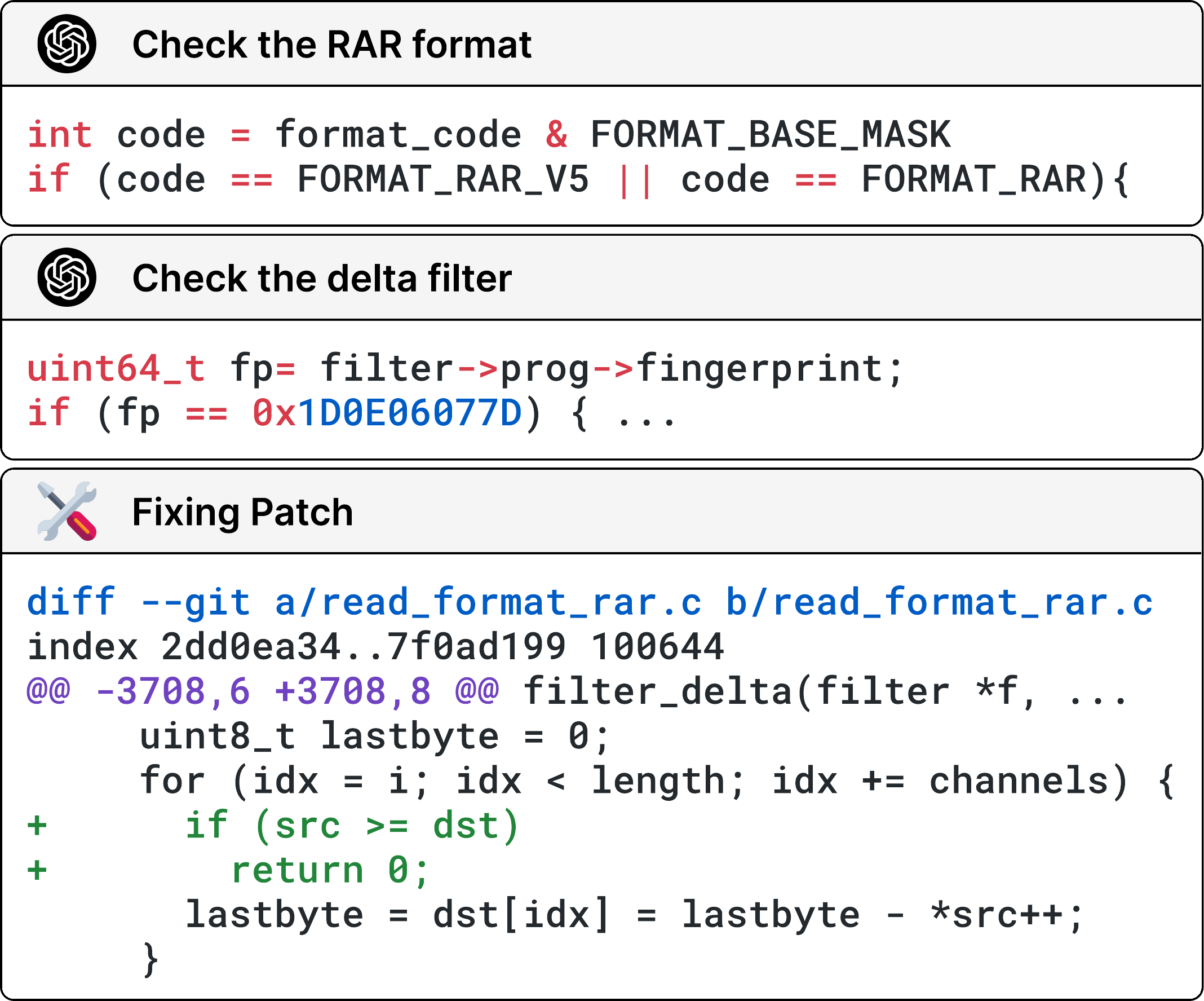}
\caption{A previously unknown vulnerability in \texttt{libarchive}}
\label{fig:real-world-example}
\end{figure}

\Cref{fig:real-world-example} shows the predicates synthesized by \pname that check the input is a valid RAR format file and that the delta filter processing logic will be triggered during decompression.
A predicate comparing the addresses of \texttt{src} and \texttt{dst} seems more helpful, but these pointers are set and computed inside the vulnerable function.
By the time execution reaches it, the predicate is too late to guide the fuzzer effectively.

\para{Apply \pname's design to other languages}
\pname analyzes program behaviors using LLMs, which have been shown to have strong performance on code analysis tasks across various programming languages \cite{jimenez2024swe-bench,ding2024primevul, gu2024cruxeval}.
In this section, we include preliminary results on extending Locus to broader programming languages.
We replace the synthesizer agent with Claude Code \cite{claude-code} to ensure greater compatibility.
Since KLEE offers limited support for non-C languages, we employ a self-reflection strategy that enables the agent to validate the semantic correctness of generated predicates.

Due to the limited availability of directed fuzzers for other languages, we apply \pname to three coverage-guided fuzzers in other ecosystems: Go-fuzz \cite{go-fuzz} for Go, Jazzer \cite{jazzer} for Java, and Atheris \cite{atheris} for Python.
For each language, we select two publicly reported vulnerabilities and use LLMs to generate canaries from the corresponding fixing patches (consistent with \Cref{subsec:case_study}).
We use Claude Sonnet 4.5 as the agent model.

The preliminary results (shown in \Cref{tab:extend}) demonstrate that \pname is able to generalize across different languages.
For Python and Go, \pname achieves an average TTE speedup of 2.6\(\times\) and 4.5\(\times\), respectively.
For the two Java vulnerabilities, while \pname does not trigger the target vulnerability in the given time budget, we observe that the vulnerable function is covered, whereas it is not even reached using Jazzer solely without \pname.

\begin{table}[!t]

\caption{TTE for applying \pname to other languages}
\label{tab:extend}
\centering
\small
\setlength{\tabcolsep}{10pt} %

\begin{tabular}{llrr}
\toprule
\textbf{Vulnerability} & \textbf{Language} & \textbf{Baseline} & \textbf{\pname} \\
\midrule
idna\#108       & Python & 14   & \textbf{5}  \\
CVE-2020-35655  & Python & T.O. & \textbf{36509} \\
hjson\#17       & Go & 18354 & \textbf{2298}  \\
gojay\#32       & Go & T.O. & \textbf{74732} \\
CVE-2021-44228  & Java  & T.O. & T.O. \\
CVE-2021-37714  & Java  & T.O. & T.O. \\
\bottomrule
\end{tabular}
\end{table}

\section{Related Work}

\para{Directed grey-box fuzzing}
Directed grey-box fuzzing is challenging, primarily due to the prohibitively large search space with sparse rewards\cite{li2024sdfuzz, luo2023selectfuzz}, \ie unlike coverage-guided fuzzing where any new coverage indicates progress.
Most prior works develop heuristics through static analysis by computing distances to target states to narrow down the search space \cite{she2024fox, shah2022mc2, huang2024titan, bohme2017aflgo, luo2023selectfuzz, srivastava2022one, li2024sdfuzz, chen2018angora, lee2021cafl}.
For example, AFLGo\cite{bohme2017aflgo} uses distance metrics between test inputs and target basic blocks to prioritize seeds that are closer to the target, while CAFL\cite{lee2021cafl} further improves this metric by introducing specialized progress-capturing state representation and calculating the distance from the testing inputs to the nearest state.

Similar to \pname, some recent approaches\cite{meng2022ltl-fuzzer, peng2018t-fuzz, ba2022stateful} also explore rewriting the program to direct execution towards hard-to-reach states.
These approaches focus on checking sophisticated conditions uncovered by human experts, \eg temporal properties in network protocols, and may not generalize to broader types of target states and can also incur expensive manual effort.
In contrast, \pname offers a generalizable framework for diverse types of programs and target states with minimal human intervention.

\para{LLM-based fuzzing}
LLMs have demonstrated promising code understanding and analysis capabilities\citep{ding2024traced, wang2024llmdfa, ding2024semcoder, gong2024evaluation, gu2024cruxeval, yang2025knighter}.
There has been growing interest in applying them to support fuzzing~\citep{aixcc}. 
Most existing approaches utilize LLMs to directly generate test inputs for the target program~\citep{zhang2024ecg, hu2023augmenting, zhang2024llamafuzz, yang2023kernelgpt, yang2024whitefox, deng2024large, deng2023titanfuzz, xia2024fuzz4all} or build grammar-aware input generators (\ie fuzzing harness) to constrain the search space~\citep{liu2024oss-fuzz-gen, huang2024halo, liu2024evaluating, shi2024harnessing, zhang2024effective}.
While these methods demonstrate the potential of LLMs to accelerate fuzzing, they often require LLMs to reason about the target program states \emph{all the way} to the inputs, making them susceptible to fundamental challenges in program analysis, such as path explosions and alias dependencies~\citep{jiang2024towards, ding2024vulnerability, li2024llm, risse2024uncovering}, while also pose challenges to LLMs in terms of the bloated context length and unbounded errors. 

As opposed to solely generating inputs or fuzzing harnesses, \pname advances LLM-based fuzzing by constraining its generation at the level of progress-capturing predicates.
Such a task formulation enables the LLM to operate within the (relatively) local context, allowing \pname to detect certain code behaviors that only emerge midway through execution to inform subsequent input searches, and also facilitates the rigorous validation of potential LLM errors.
As a result, this strategy can effectively sidestep reasoning across lengthy function call chains, enabling a greater focus on more local, intermediate function contexts.

\para{LLM-driven proof synthesis}
\pname shares a similar spirit with recent research that focuses on reasoning about loop invariants and program specifications to synthesize proofs and automate program verification\cite{chakraborty2023ranking, chen2024automated, lu2024proof, kozyrev2024coqpilot}.
\pname resembles these approaches in the sense that they also integrate LLMs with rule-based verifiers, such as theorem provers or SMT solvers, to check the validity of synthesized properties.
However, \pname relaxes the requirement that the synthesized specifications must facilitate rigorous proof steps, but treats the predicates as best-effort guidance for input search.
It bridges the gap between formal verification and dynamic testing, leveraging the reasoning capabilities of LLMs to identify meaningful program states while prioritizing efficiency and generality over formal guarantees.

\section{Threats to Validity} 
\label{sec:threat2valid}

\para{Threats to predicate validation}
While \pname leverages symbolic execution to bound the error of the synthesizer, it inherits all of its limitations. 
For example, we still face the path explosion problem, especially when there are loops between the generated predicates and the canary. 
While we adopt the common strategies to address these limitations, \eg loop unrolling, we cannot formally guarantee that the relaxation brought by the generated predicates is always valid and can thus effectively guide fuzzing.
This leads to orthogonal but interesting future directions on loop invariant reasoning and function summary (using LLMs) for symbolic execution.

\para{Threats to LLM types}
Second, our evaluation covers only a limited set of LLMs and benchmark programs.
The generality of our results to other models, especially open-weight LLMs and broader software systems, remains to be validated. 
The performance of our approach may depend on the underlying LLM capability and training data. 
Additionally, our evaluated benchmark Magma, though built on popular software and real-world vulnerabilities, may not capture the comprehensive challenges present in diverse software.

\para{Threats to data leakage}
While the task formulation, \ie predicate synthesis for directed fuzzing, is arguably hard to suffer from the data leakage problem, the software project in our evaluation is largely included in the training data of any major LLMs.
As the ultimate goal is to detect and confirm vulnerabilities, we believe that such a threat is minor compared to the practical security impact, as we have shown in \Cref{subsec:case_study}.

\section{Conclusion}

This paper presented \pname, a novel framework for enhancing directed fuzzing through synthetic progress-capturing predicates. 
With an agentic synthesizer-validator architecture, \pname effectively guides fuzzing via the predicates while also ensuring any errors in the synthesized predicates are fixed by symbolic execution. 
Our evaluation demonstrated that \pname substantially improves the state-of-the-art fuzzers.
So far, it has uncovered nine previously unpatched vulnerabilities across three software projects, with three already acknowledged with draft fixes.

\begin{acks}
We are grateful to Jun Yang, Weichen Li, Chenghao Yang, Chenyuan Yang, Zheng Yu, Heqing Huang, Penghui Li, Weiteng Chen, and Miltos Allamanis for sharing their constructive and insightful feedback, which significantly helped improve this paper.
This research is supported in part by Open Philanthropy, NSF CAREER Award CNS-2442719, and generous gifts from OpenAI.
Any opinions, findings, and conclusions or recommendations expressed
in this material are those of the author(s) and do not necessarily
reflect the views of the sponsors.
Results presented in this paper were obtained using the Chameleon testbed supported by the National Science Foundation.
Language model assistants were used to help polish the paper.
\end{acks}

\bibliographystyle{ACM-Reference-Format}
\bibliography{zotero, extra}

\end{document}